\documentclass[useAMS,usenatbib]{mn2e}

\usepackage{graphicx}

\title[Determining the orbital parameters of WR30a]
{Modeling the line variations from the wind-wind shock emissions of WR 30a}
\author[D. Falceta-Gon\c{c}alves, Z. Abraham \& V. Jatenco-Pereira]
{D. Falceta-Gon\c{c}alves$^{1}$\thanks{E-mail:diego.goncalves@unicsul.br}, 
Z. Abraham$^{2}$ and V. Jatenco-Pereira$^{2}$ \\
$^{1}$N\' ucleo de Astrof\' isica Te\' orica, CETEC - Universidade Cruzeiro do 
Sul, Rua Galv\~ ao Bueno 868, 01506-000, S\~ao Paulo, Brazil \\
$^{2}$Instituto de Astronomia, Geof\'\i sica e Ci\^encias Atmosf\'ericas, Universidade de S\~ao Paulo, 
01060-970, S\~ao Paulo, Brazil }

\begin{document}

\date{ }

\pagerange{\pageref{firstpage}--\pageref{lastpage}} \pubyear{2007}

\maketitle

\label{firstpage}

\begin{abstract}

The study of Wolf-Rayet stars plays an important role in evolutionary theories of massive stars. Among these objects, $\sim 20\%$ are known to 
be in binary systems and can therefore be used for the mass determination of these stars. Most of these systems are not spatially resolved and spectral lines 
can be used to constrain the orbital parameters. However, part of the emission may originate in the interaction 
zone between the stellar winds, 
modifying the line profiles and thus challenging us to use different models to interpret them.  
In this work, we analyzed the He{\sevensize II}$\lambda$4686\AA \ + C{\sevensize IV}$\lambda$4658\AA \ blended lines of WR30a (WO4+O5) 
assuming that part of the emission originate in the wind-wind interaction zone. In fact, this 
line presents a quiescent base profile, attributed to the  WO wind, and a superposed excess, which 
varies with  the orbital phase along the  4.6 day period. Under these assumptions, we were able to fit 
the excess spectral line profile and central velocity for all phases, except for the longest wavelengths, 
where a spectral line with constant velocity seems to be present. The fit parameters provide the 
eccentricity and inclination of the binary orbit, from which it is possible to constrain the stellar masses.

\end{abstract}


\section{Introduction}

Massive stars are known to drive strong winds, which are responsible for the transfer 
of a large amount of the stellar mass to the interstellar medium, contributing to the 
feedback of 
chemical elements, and to the creation of  cloud cavities in which these objects are found. 
Typically, O 
stars present mass-loss rates of $\dot{M}_{\rm O} \sim 10^{-6} - 10^{-5}$ M$_\odot$ yr$^{-1}$ and wind velocities 
of $v_{\rm O} \sim 2000 - 3500$ km s$^{-1}$, while Wolf-Rayet (WR) stars present $\dot{M}_{\rm WR} \sim 10^{-6} 
- 10^{-4}$ M$_\odot$ yr$^{-1}$ and $v_{\rm WR} \sim 1000 - 4000$ km s$^{-1}$ (Nugis \& Lamers 2000, Lamers 2001).

In  massive binary systems in which  both stars present high mass-loss rates and
high-velocity winds, the collision of the winds will occur. Therefore, a contact surface is formed where the 
momenta of the two winds are equal, surrounded by two shocks. The post-shocked gas, cools as it flows along 
the contact surface and is responsible for strong free-free emission at X-ray and radio wavelengths.  
X-rays in massive binary systems are orders of magnitude higher than those observed in single massive stars, and 
are used as an indication of binarity. 

UV and optical lines can also indicate the binary nature of a given object, when they present  
periodic profile variations.  If the lines are of photospheric or atmospheric origin, as the stars 
move along their orbit, they  suffer Doppler shifts, which   depend on the orbital phase and 
inclination. However, some massive objects show periodic variable line profiles that cannot be explained 
under these assumptions. 

Seggewiss (1974) noted that the binary system WR79 presented two peaks, 
superimposed to the C{\sevensize III} emission line, which changed their position and intensity with time. 
Typically, in double line spectroscopic binaries, each peak moves in a different 
direction, indicating opposite velocity components along the 
line of sight for each star; however in WR79 both peaks moved in the same direction. 
L$\rm \ddot{u}$hrs (1997) presented a model in which the two peaks were not produced by the 
stellar photosphere, but were generated by the flowing gas at the contact surface 
between the two strong shocks. This model reproduced well the data for WR 
79, but failed to reproduce the line profiles of other WR binary systems, among them WR 30a (Bartzakos, Moffat 
\& Niemela 2001). Falceta-Gon\c calves, Abraham \& Jatenco-Pereira (2006) improved  L$\rm \ddot{u}$hrs' model introducing more realistic parameters, as stream turbulence and  gas opacity, to account for the line broadening and peak displacement.  

In the present work, we applied this model to WR 30a, which was classified as a WO4+O5 binary system 
(Moffat \& Seggewiss 1984, Crowther et al. 1998); its binary nature was reported by Niemela (1995) 
based on spectral-line radial  velocities obtained with a high temporal resolution. 
Later, Gosset et al. (2001) presented a detailed analysis of the spectra of WR30a 
for several epochs and were able to confirm the binary hypothesis and to determine the period of $P \sim 4.6$ days. 
They also noted strong line-profile variations, which made more difficult  the determination of the 
stellar mass ratio and the orbital inclination from standard methods. They concluded that the C{\sevensize IV}$\lambda$4658\AA \ 
line-profile variations were related to wind-wind collision processes, but did not model the lines under such an 
assumption. The same conclusion was reached by Bartzakos, Moffat \& Niemela (2001) 
using the C{\sevensize IV}$\lambda$5801\AA \ line. Also, Paardekooper et al. (2003) presented photometric 
measurements at $V$ and $B$ bands; the light curves confirmed the period obtained by Gosset et al. (2001), but 
also showed higher frequency variability in the $V$ band, with timescale  of hours. 
They concluded that this could be due to the strong variability of the C{\sevensize IV}$\lambda$5801\AA \ line, 
possibly related to the wind-wind interaction. 

In the present work, we tested the wind-wind shock emission hypothesis on the C{\sevensize IV}$\lambda$4658 \AA \ excess line profile variations measured by Gosset et al. (2001) using the 
model developed by Falceta-Gon\c calves, Abraham \& Jatenco-Pereira (2006), which is briefly described in Section 2. In Section 3, we show the results obtained for WR 30a and present a brief discussion, followed by the conclusions in Section 4.


\section{Wind-wind emission model}

In the proposed situation, in which both stars present high mass-loss rates in supersonic winds, 
the contact surface, which is schematically shown in Figure 1, will have a geometry described analytically 
by (Luo, McCray \& Mac-Low 1990):

\begin{equation}
\medskip
\frac{dy}{dz}= \frac {(\eta^{-1/2}{d_2}^2+ {d_1}^2)y}{\eta^{-1/2}{d_2}^2z+{d_1}^2(z-D)},
\medskip
\end{equation}

\noindent
where $D$ is the distance between the  stars; $d_1$ and $d_2$ are the distances of the primary and secondary 
stars to the contact surface, respectively, and $\eta=\dot{M_s} v_s/\dot{M_p} v_p$, where  $\dot{M_p}$ and 
$\dot{M_s}$ are the mass-loss rates of the primary and the secondary stars, and $v_p$ and $v_s$ their respective wind velocities. 
The contact surface will asymptotically have a conical shape with an opening angle defined by $\beta$, given 
by:

\begin{equation}
\medskip
\beta \simeq 120^{\circ} \left(1-\frac{\eta^{2/5}}{4} \right) \eta^{1/3},
\medskip
\end{equation}

\noindent
and the apex will occur at a distance to the primary star given by:

\begin{equation}
\medskip
d_{1} |_{\rm apex} = D / (1+\eta^{1/2}).
\medskip
\end{equation}

Two shock fronts will be formed on both sides of the contact surface, generated by each wind, and the 
gas in the post-shock region will flow away along the contact surface. 
While  it flows, the gas will cool due to expansion and radiation. This emission, detectable from 
radio wavelengths to X-rays, is the signature of wind-wind collisions (Usov 1992, Falceta-Gon\c calves, 
Jatenco-Pereira \& Abraham 2005, Abraham et al. 2005, Pittard \& Dougherty 2006). The stream of hot gas 
flowing along the contact surface will, eventually, reach temperatures that allow the recombination of certain elements. 
In this situation, the observer will detect the emission line shifted due to the stream velocity component 
along the line of sight, which may intercept regions with different velocity components, as shown in Figure 1. 
The shaded regions in the shock layer represent the emitting fluid elements. 
Here,  observer $A$ would detect two emission lines,  a blueshifted line from fluid element 1, and a 
redshifted line from element 2. 
On the other hand, observer $B$ would detect two blueshifted lines. 

\begin{figure}
\centering
\includegraphics[width=8cm]{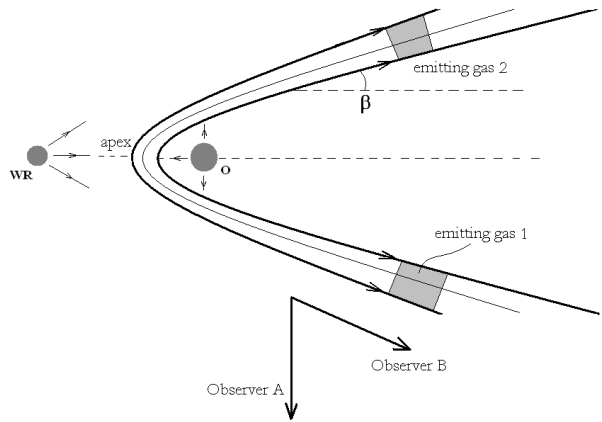}
\caption {Schematic view of the wind-wind interaction surface and the model geometry.}
\label{figure2}
\end{figure}

L$\rm \ddot{u}$hrs (1997) used these concepts to model  line-profile variability during the orbital period of 
WR79. To reproduce the observations, he assumed a large emission region, limited by two cones with  aperture 
angles $\beta$ and $\beta + \Delta \beta$, selected $ad$ $hoc$ to fit the the observations.   However, the 
assumption is  valid only in the case of quasi-adiabatic shocks, which occur in long-period systems. 
Radiative shocks evolve much faster, creating layers that are narrow and turbulent (Stevens, Blondin \& 
Pollock 1992), and therefore cannot simply be described by a fixed range of beta values as assumed 
by L$\rm \ddot{u}$hrs (1997). Falceta-Gon\c calves, 
Abraham \& Jatenco-Pereira (2006) by-passed this problem by including turbulence in the stream layer, and also 
took into account the shocked gas opacity, which may result in line profile asymmetries. 
In that work, the line profile is obtained by integrating the emission intensities of each fluid element as:

\begin{equation}
\medskip
I(v) = \mathcal{C(\varphi)} \int_{0}^{\pi }\exp \left[ -\frac{\left( v-v_{\rm obs}\right) ^{2}}{2\sigma
^{2}}\right]e^{-\tau(\alpha)} d\alpha, 
\medskip
\end{equation}

\noindent
where:

\begin{eqnarray}
\medskip
v_{\rm obs} = v_{\rm flow} (-\cos \beta \cos \varphi\sin i + \sin \beta \cos 
\alpha \sin \varphi \sin i \nonumber \\
- \sin \beta \sin \alpha \cos i),
\medskip
\end{eqnarray} 

\noindent
is the observed stream velocity component projected into the line of sight, $\alpha$ is the  azimuthal 
angle of the shock cone, $i$ is the orbital inclination, $\varphi$ is the orbital phase angle, $\sigma$ 
is the turbulent velocity amplitude, $\tau$ is the optical depth along the line of sight and $\mathcal{C(\varphi)}$ 
is a normalization constant. 

\begin{figure}
\centering
\includegraphics[width=6cm]{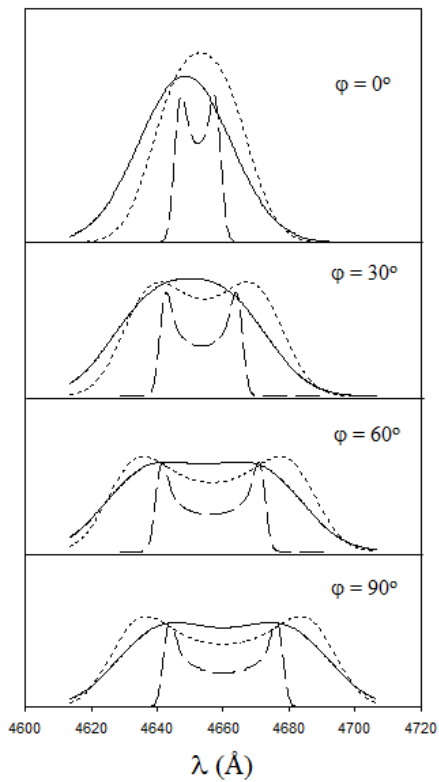}
\caption {Synthetic line profiles for models (a): $\beta = 55\degr$, $i = 60\degr$, 
$\sigma = 0.1$ and $v_{\rm flow} = 1000$ km s$^{-1}$ (dashed line) (b): $\beta = 55\degr$, 
$i = 60\degr$, 
$\sigma = 0.5$ and $v_{\rm flow} = 1500$ km s$^{-1}$ (solid line) and (c): $\beta = 
70\degr$, $i = 60\degr$, 
$\sigma = 0.3$ and $v_{\rm flow} = 1000$ km s$^{-1}$ (dotted line); for different orbital 
phases ranging from $\varphi = 0\degr$ to $90\degr$.}
\label{figure2}
\end{figure}

In Figure 2, we show the model results for different orbital and stream  parameters. The dashed line model (a) 
was calculated using $\beta = 55^\circ$, $i = 60^\circ$, $\sigma = 0.1$ and $v_{\rm flow} = 1000$ km s$^{-1}$; 
the solid line (b) using $\beta = 55^\circ$, $i = 60^\circ$, $\sigma = 0.5$ and $v_{\rm flow} = 1500$ km s$^{-1}$; 
and the dotted line (c) using $\beta = 70^\circ$, $i = 60^\circ$, $\sigma = 0.3$ and $v_{\rm flow} = 1000$ km s$^{-1}$. 
It is noticeable that, e.g. in models (a) and (c), there are two peaks generated by each of the shock-cone layers 
intercepted by the line of sight. As shown by  Falceta-Gon\c calves et al. (2006), this is valid for 
$0\degr \leq i < 90\degr$, while  for $i = 90^{\circ}$, the profile will present a single peak at 
$\varphi = 0\degr$, which corresponds to the cone axis coincident with the line of sight. Another comparative 
analysis between (a) and (c) shows the effect of changing $\beta$; as  $\beta$ increases, the distance between 
the two peaks becomes larger. Models (b) and (c) present high turbulence amplitude, which leads to a larger 
line broadening. Here, we neglected the gas opacity (i.e. $\tau << 1$) to show the presence of two peaks in 
the line profiles. If the shock cone is optically thick, part of the emission along the line of sight will 
be absorbed, and the redder peak will be less intense or, eventually, completely disappear.   

In the next section we present the application of this model to  the C{\sevensize IV}$\lambda$4658\AA \ 
excess line-profiles of 
WR30a, in order  to determine the orbital and stellar parameters.


\section{The case of WR30a}

Gosset et al. (2001) presented a detailed spectroscopic study of WR30a with data covering 30 days, which corresponds to 
about 6 complete orbital cycles. From their observations, the strongest broad emission line is identified as 
a blend dominated by C{\sevensize IV}$\lambda$4658\AA \ and He{\sevensize II}$\lambda$4686\AA . Subtracting a constant 
parabolic base line, attributed 
to the WR stellar wind emission, they obtained the averaged residual spectra shown in Figure 3. 
This excess emission presents anomalous profiles and variability in both line intensity and central wavelenght, 
all related to the orbital phase. 
This effect could originate in selective wind eclipses, resulting in different profiles as the O-star is located behind 
or in 
front of the WR star along the orbital motion. However, this is not the case in WR30a because due to its low orbital 
inclination no phase-dependent atmospheric absorption would be expected.

\begin{figure}
\centering
\includegraphics[width=6cm]{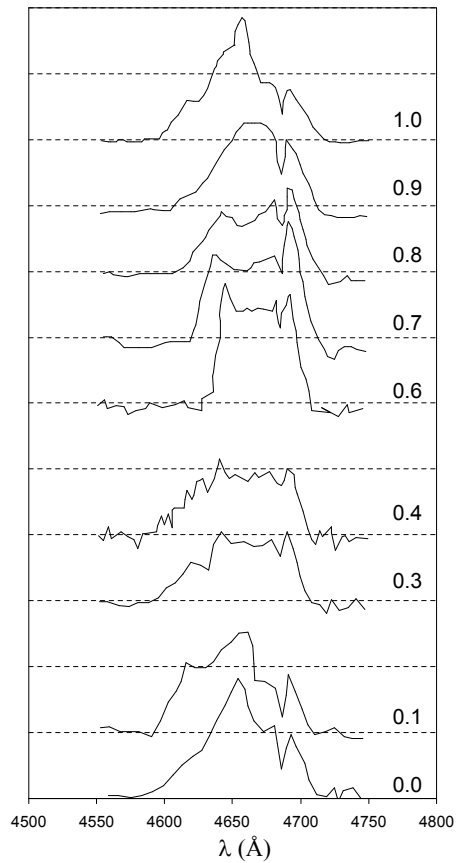}
\caption {Excess emission line profiles obtained from Gosset $et$ $al.$ (2001).}
\label{figure2}
\end{figure}

\subsection{The orbital parameters}

As it is clear from Figure 2, a wind-wind interaction model is compatible with the line profiles observed in WR30a. Hence, we applied the model described in the 
previous section to determine the binary-system parameters. An important model parameter in equation (4) is the optical depth $\tau(\alpha)$ of the absorbing material; 
Falceta-Gon\c calves et al. (2006) assumed that it is the absorption by the dense material that accumulates along the contact surface that produces the asymmetries in 
the line profiles, while any other  absorption is taken into account by the multiplying  factor  $C(\varphi)$. Since most of the lines seem to be symmetric, at least within the 
uncertainties due to the subtraction of a phase independent base profile, we assumed  $\tau(\alpha) = 0$. It turned out from our fitting that $C(\varphi)$ is also independent
 of the orbital phase..

\begin{figure}
\centering
\includegraphics[width=6cm]{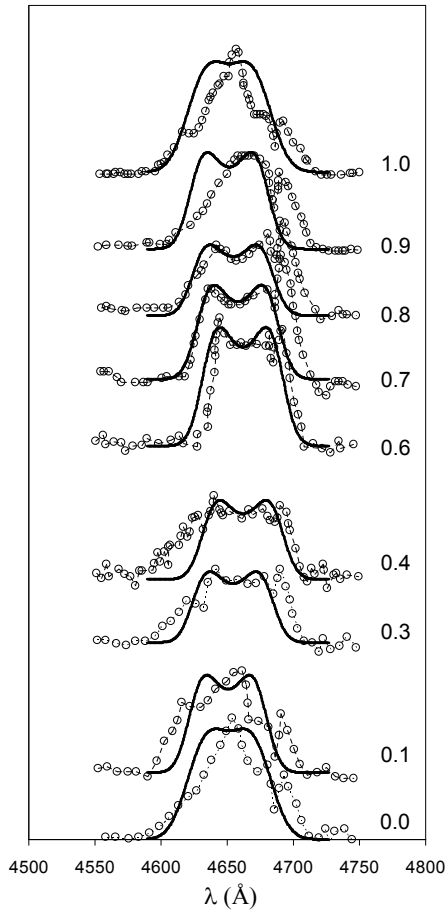}
\caption {Best fitting synthetic line profiles obtained for $\beta = 50^\circ$, $i = 
20^\circ$, $\sigma = 0.3$ and $v_{flow} = 2200$km s$^{-1}$. The orbital phase of each data set is exactly the 
same shown in Figure 2.}
\label{figure2}
\end{figure}

In order to model the observed excess emission spectra  for each orbital phase, we performed calculations 
varying the opening angle, the turbulence amplitude and the stream velocity for different values of the orbital 
inclination.  
We were not able to fit simultaneously the complete width of the observed profiles and their mean velocities 
for all epochs, even for high values of the orbital inclination. Instead, we obtained a better agreement when 
we considered that the  emission at the longest wavelengths, seen as a peak near 4690\AA \, 
is not formed in the  shock region.
The best fit occurred for $\beta = 50^\circ \pm 5\degr$, $i = 20^\circ \pm 5\degr$, $v_{\rm flow} = 2200 \pm 
500$ km s$^{-1}$, and $\sigma = 0.3 \pm 0.1$; the corresponding profiles  are shown in Figure 4. The  phase 
angles used for the fit for each epoch, and their respective uncertainties are shown in Figure 5, as a 
function of the orbital phase.
The uncertainties in the parameters were obtained by changing each one of them independently until the model 
became incompatible with the observations.  The excess profile for orbital phase 0.7 is very well-fitted by 
the wind-wind emission model, with the exception of the already mentioned  additional peak near 4690\AA. 
For orbital phases near 1.0  the model profiles are wider than those observed, which seemed to be single peaked. 
This may be related to the subtracted base line profile, which was obtained by Gosset et al. (2001) for phase 
0.55 and equally applied to all orbital phases. 

If the subtracted profile is produced in the WR stellar wind, as it seems to be the case, it would be necessary to take into account the fact that the wind is not completely 
symmetric but shaped by the conic hole with aperture $\beta$, as the wind is deflected at the shock. Since the orientation of this hole changes with orbital phase, 
we would expect also changes in the broad profile produced by the WR wind. For the particular case of $i=90^{\circ}$, when the WR star is in conjunction 
regarding the observer, there will be a stronger deficit of emission at the redshifted part of the line profile. On the other hand, if the WR star is in opposition, 
the amount of upcoming gas will be lower, affecting more the blueshifted emission. For $i=0^{\circ}$, the profile would not change as the orientation of the 
cone is always the same with respect to the observer. For WR30a, with $i \simeq 20^{\circ}$, at WR conjunction there will be mainly an effect on the 
red side of the profile but the blues side would also be affected. The opposite would occur at WR opposition. This is compatible with the narrower excess profiles near phase 
1.0 as can be seen in Figure 4.

At  long wavelengths, the excess near 4690\AA \ could possibly be the blended emission from the 
He{\sevensize II}$\lambda$4686\AA \ 
line; although it seems decoupled from the shock emission during the orbital period, it is too narrow $\sim 
20$\AA \ (1300 km s$^{-1}$) to be attributed to the WR stellar wind ($> 2000$km s$^{-1}$). It could be, however, 
a fraction of the WR P-Cygni profile or it could also  be related to the 
remnant shocked material surrounding the stellar system, ionized by the stellar radiation. Gosset et al. (2001) 
also attributed the narrow absorption 
superposed on the broad excess emission to a He{\sevensize II}$\lambda$4686\AA \ absorption line from the 
O star.

The  phase angles derived from our model provide additional information regarding the orbit, namely its 
eccentricity, since the model strictly depends on the angle between the cone axis and the line of the sight, 
while the observed profiles depend on the orbital phase. The dependence between the two parameters is related 
to the orbital eccentricity, as shown in Figure 5 for $e = 0.0, 0.2$ and $0.5$. 
The dotted line represent the result for $e = 0.0$ (circular orbit), in which the phase angle changes linearly 
with time. The dashed line shows the fast 
phase angle variation during the periastron passage for $e = 0.5$. The data for WR30a 
seem to fit best the curve corresponding to $e = 0.2$ when the lines are shifted by $\varphi_0 = -20^{\circ}$ 
in phase angle (0.056 of the orbital phase). Part of this shift could by explained  by the   {\it Coriolis Effect},  
which produces a deviation of the cone axis as a consequence of the relative motion of the stars; however, 
Gosset et. al. (2001) estimated this effect as only 0.01 in orbital phase. Most of the contribution is probably 
produced by the binning  of the data in intervals of one tenth of the period, which could also explain why the  
eccentricity we found is different from zero.

\begin{figure}
\centering
\includegraphics[width=8cm]{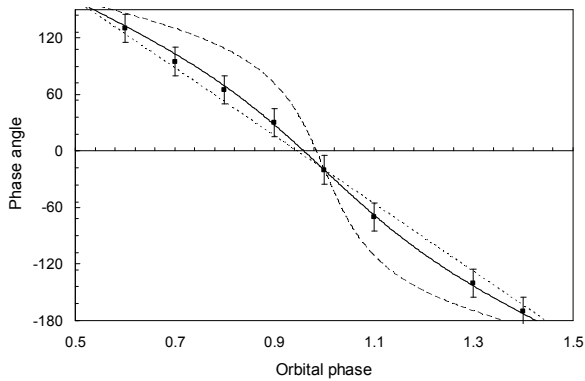}
\caption {The dependence of the phase angle on the orbital phase for  
eccentricities e = 0.0 (dotted), 0.2 (solid) and 0.5 (dashed), where 
$\varphi_0 = -20^\circ$. The data for WR30a (squares) were obtained from the 
phase angles of the fits and the observed orbital phases of Figure 3.}
\label{figure5}
\end{figure}

\subsection{The stellar masses}

After modeling the wind-wind shock emission, and determining the orbital parameters of the binary system, we can 
constrain the stellar masses. For this purpose, we analyzed the velocity curve of the O-star obtained by Gosset 
et al. (2001)  from measurements of four lines (He{\sevensize II}$\lambda$4542\AA, H$\gamma$, 
He{\sevensize II}$\lambda$4200\AA \ and H8) during the orbital period. Their results are shown in Figure 6, as well as the fit of the modeled 
curve for $e = 0.2$ and $K_{\rm O} = 25$ km s$^{-1}$. As a result, the mass function: 

\begin{equation}
\medskip
f(m) = \frac{K_{\rm O}^3 P}{2\pi G} = \frac{(M_{\rm WR} \sin i)^3}{(M_{\rm WR}+M_{\rm O})^2} ,
\medskip
\end{equation}

\noindent
gives $f(m) = 0.0074$ M$_{\odot}$. If we assume $M_{\rm O} = 40$ 
M$_{\odot}$ (60 M$_{\odot}$), using $i = 20\degr$ obtained from the shock emission, we 
find $M_{\rm WR} = 7.5$ M$_{\odot}$ (9.7 M$_{\odot}$), which gives a mass ratio 
$M_{\rm WR} / M_{\rm O} \sim 0.2$. 
Regarding the orbit, using the obtained mass function and orbital 
inclination, we find $a_{\rm O} \simeq 5.4$ R$_{\odot}$ and 
$a_{\rm WR} \simeq 30$ R$_{\odot}$.

These  results are in  agreement with those  expected for the masses of O and WO stars  $40-60$ and $8-10$ M$_{\odot}$, 
respectively (Howarth \& Prinja 1989, de Marco \& Schmutz 1999); also, the distance between the stars  is larger than 
the sum of the expected stellar radii: $D > R_{\rm O} + R_{\rm WR} \sim 20$ R$_{\odot}$ (Schaerer, Schmutz \& Grenon 
1997). 

From Equation 2 and the value  of the cone-opening angle ($\beta = 50^\circ$) we calculated the wind momentum ratio 
and obtained $\eta = 0.12$, which is compatible with typical observed values in systems similar to WR30a (Lamers 2001).
From Equation 3, we  estimated the  distance from the apex to the stars: for $\eta = 0.12$ and the total separation 
$D = 35.4$ R$_{\odot}$ we obtained  $d_{1} |_{\rm apex} ~ 25$ R$_{\odot}$ and $d_{2} |_{\rm apex} ~ 10$ R$_{\odot}$. 
Considering a typical O-star, $d_{2} |_{\rm apex}$ would be of the same order than the stellar radius, and  the 
frontal part of its wind could be crushed onto the stellar surface. The shape of the contact surface would be 
spherical near the apex,  affecting maybe the X-ray emission from the system (Usov 1992), but our model will still 
be valid, since the line emission originates at larger distances from the apex, on the contact surface. 

\begin{figure}
\centering
\includegraphics[width=8cm]{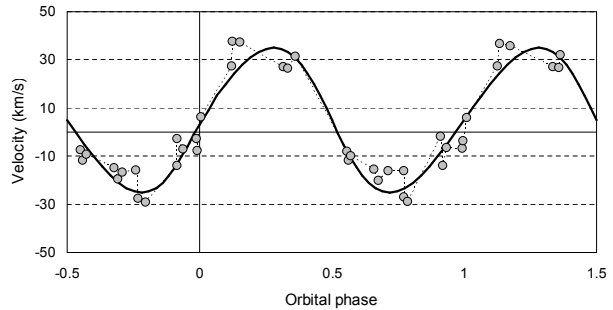}
\caption {The O-star radial velocity as function of the orbital phase. The observed data (circles) 
were obtained from Gosset et al. (2001). The solid line represent the fit for $K_{O} = 25$ 
km s$^{-1}$ and $e = 0.2$.}
\label{figure2}
\end{figure}


\section{Conclusions}

In this work we presented an application of the wind-wind shock emission model proposed by Falceta-Gon\c calves et al. (2006) to
the line profile variations of WR 30a. In this model, the wind-wind shock structure can be represented by 
a cone, along which the shocked material flows. This gas will emit radiation, cool down  and eventually 
reach recombination temperatures. The observed emission lines will then suffer Doppler shifts
due to the stream velocity component along the line of sight. During the orbital movement, the 
cone position will change, as well as the radial velocity, which will cause the line profile variations 
observed in several massive binary systems. 

Gosset et al. (2001) obtained detailed spectra of WR30a during more than 30 days. They determined 
the orbital period ($P = 4.6$d), and obtained the radial velocity curve for the O-star. Regarding 
the WR component, they found that the blended He{\sevensize II}$\lambda$4686\AA \ and C{\sevensize IV}$\lambda$4658\AA \ lines showed a variable 
excess emission. 
In the present paper we modeled this variable emission, being able to reproduce the variations except for the red 
part of the profiles, which seemed to be unchanged in velocity and were probably generated in the stellar wind of 
the WR star instead of in the shock region.

The best-fitting result was obtained for  $\beta = 50^\circ$, $i = 
20^\circ$, $\sigma = 0.3$ and $v_{\rm flow} = 2200$km s$^{-1}$. Also, correlating the orbital 
phase with the modeled phase angle, it was possible to determine the orbital eccentricity as
$e = 0.2$, similar to the value of 0.0 previously assumed {\it ad hoc} by other authors. Although both  values lead 
to very  small differences in the orbital shape, its value is important for the determination of the stellar masses 
and orbital separation between the stars. Using this eccentricity and orbital inclination to model the radial 
velocity curve 
for the O-star, we found $K_{\rm O} = 25$ km s$^{-1}$. If we assume $M_{\rm O} = 40 - 60$ 
M$_{\odot}$, we find $M_{\rm WR} = 7.5 - 9.7$ M$_{\odot}$, and the orbital major semi-axis $a_{\rm O} \simeq 5.4$ 
R$_{\odot}$ and 
$a_{\rm WR} \simeq 30$ R$_{\odot}$.

The model showed itself a powerful tool for constraining the wind and orbital parameters of 
massive binary systems.  The fits did not matched the data exactly at all epochs, but considering the difficulties 
of subtracting the emission excess from the original spectra, the general shape 
and peak position variations were well reproduced. 
We must state that this is a very simple approximation, taking into account the complexities 
found in such systems. Numerical simulations could give a more detailed analysis, and probably more 
accurate values for the model parameters in future works.


\section*{Acknowledgments}

      D.F.G thanks FAPESP (No. 06/57824-1 and 07/50065-0) for financial support. Z.A. and V.J.P. thanks 
FAPESP, CNPq and FINEP for support. We also thank Nicole St-Louis for the useful comments 
and suggestions to improve this work.

\end{document}